\documentclass[12pt,preprint]{aastex}

\shorttitle{Radio polarimetry of strong turbulence}
\shortauthors{Stroman \& Pohl}
\usepackage{natbib}

\begin{document}

\title{Radio polarimetry signatures of strong magnetic turbulence in
Supernova Remnants}

\author{Wendy Stroman}
\affil{Department of Physics and Astronomy, Iowa State University, Ames,
IA 50011}
\and
\author{Martin Pohl}
\affil{Department of Physics and Astronomy, Iowa State University, Ames,
IA 50011}
\email{mkp@iastate.edu}
\begin{abstract}
We discuss the emission and transport of polarized radio-band synchrotron radiation near
the forward shocks of young shell-type supernova remnants, for which X-ray data
indicate a strong amplification of turbulent magnetic field. Modeling the magnetic turbulence 
through the superposition of waves, we calculate the degree of polarization and the magnetic 
polarization direction which is at $90^\circ$ to the conventional electric polarization 
direction.
We find that isotropic strong turbulence will produce weakly polarized radio emission
even in the absence of internal Faraday rotation. 
If anisotropy is imposed on the magnetic-field structure, the degree of polarization
can be significantly increased, provided internal Faraday rotation is inefficient. Both
for shock compression and a mixture with a homogeneous field,
the increase in polarization degree goes along with a fairly precise alignment of the
magnetic-polarization angle with the direction of
the dominant magnetic-field component, implying tangential magnetic polarization at the rims
in the case of shock compression. We compare our model with high-resolution
radio polarimetry data of Tycho's remnant. Using the absence of internal Faraday rotation
we find a soft limit for the amplitude of magnetic turbulence, $\delta B \lesssim 200\ {\rm \mu G}$.
The data are compatible with a turbulent magnetic field superimposed on a
radial large-scale field of similar amplitude, $\delta B\simeq B_0$. An alternative viable scenario
involves anisotropic turbulence with stronger amplitudes in the radial direction, as was
observed in recent MHD simulations of shocks propagating through a medium with significant
density fluctuations.\end{abstract}


\section{Introduction}
The origin of Galactic cosmic rays and the mechanisms of their acceleration are
among the most challenging problems in astrophysics.
Shell-type supernova remnants (SNR) have long been thought to be the sources of cosmic rays, 
primarily through acceleration at their powerful forward shocks.
Particle acceleration at collisionless shocks is intrinsically efficient (e.g. Kang \& Jones 2005)
and arises from pitch-angle scattering in the plasma flows that have systematically
different velocities upstream and downstream of the shock \citep{Bell78}. Detailed studies show that 
the acceleration efficiency and the resulting spectra depend on the orientation angle of 
the magnetic field and on the amplitude and characteristics of magnetic turbulence near the shock
\citep[e.g.][]{gj96,md01,g05,bt05}, part of which is self-generated. 
The amplitude of the turbulence also sets
the scale for the maximum energy to which a remnant may accelerate particles. For typical 
interstellar magnetic field values, SNRs can at best accelerate particles to $10^{15}$~eV, where
the cosmic-ray spectrum shows a break known as the {\it knee} \citep{lc83a,lc83b}. 
If the cosmic rays would drive a turbulent magnetic field to an amplitude much larger
than the homogeneous interstellar field \citep{lb00,bl01}, particle 
acceleration may be faster and extend to higher energies \citep{vladi}.  

\citet{be04} found that, rather than resonant Alfv\'en waves, the current carried by 
drifting cosmic rays should efficiently excite non-resonant, nearly purely growing 
modes of magnetic turbulence on spatial scales much smaller 
than the cosmic-ray Larmor radius. MHD simulations that assume the cosmic-ray current to be constant 
\citep{be04,be05,zira08,rev08} indeed indicate a 
strong magnetic-field amplification following an approximately isotropic plasma filamentation 
in the non-linear stage. On the other hand, recent kinetic simulations suggest that the
amplitude of the field
perturbations saturates at approximately the amplitude of the homogeneous 
upstream field on account of nonlinear backreactions \citep{nps08}. Density fluctuations in the 
upstream region will also distort the shock, leading to turbulent magnetic-field growth downstream
\citep{bal01,gj07,zp08}.

While its micro-physics is not fully understood,
magnetic-field amplification is observationally required to explain that
a large fraction of the non-thermal X-ray emission on the rims of young SNRs
is concentrated in narrow filaments \citep[e.g.][]{hug00,got01,hwa02,bam03,bam05}.
These filaments can be interpreted either as limited by rapid energy losses of the 
radiating electrons \citep{v+l03,bam03} or, alternatively, as caused by rapid damping
of strong magnetic turbulence downstream of the SNR shock \citep{pyl05}.
Both interpretations require magnetic-field amplification, and
the turbulently amplified field is expected to have a
small wavelength. A very strong magnetic field is also suggested by time-variability
of patches of non-thermal X-ray emission near the forward shock of SNR RX~J1713-3946 \citep{uch07}, 
although observational limits to the radio emission of secondary electrons \citep{h+p08}
and the cosmic-ray
e/p ratio in general \citep{k+w08} indicate that we could observe just the build-up and decay
of a magnetic structure \citep[e.g.][]{butt08,byk08}.

Because of its role in particle acceleration it is of paramount importance to understand 
the properties of strong magnetic turbulence near the forward shocks of SNRs. 
A significant uncertainty in our interpretations arises from the unknown 
spatial distribution of the strong, turbulent magnetic field. The properties of the turbulent
magnetic field can be traced through synchrotron intensity fluctuations \cite[e.g.][]{cl02}
or through polarimetry. Here we present a search for signatures of 
such turbulence in polarized radio synchrotron emission. For that purpose we create
parametrized models of magnetic turbulence and calculate the emission and transport of
linearly polarized synchrotron radiation through the turbulent downstream medium, including
the effects of Faraday rotation. The results are compared with published radio
polarimetry data of Kepler's SNR (SN~1604) \citep{del02} and Tycho's SNR (SN~1572) \citep{di91}.

Since the degree of linear polarization and the polarization angle depend mainly on the
structure of the 
turbulent magnetic field, our results are very generic and can be widely applied.
The amplitude of the magnetic field enters only through a scaling parameter
that describes the extent of depolarization through Faraday rotation.

\section{The model}
\subsection{Parametrization of the turbulent magnetic field}\label{mf}
A turbulent magnetic field can be constructed via superposition of many 
magnetic waves with random orientation.
A single magnetic wave carries a magnetic-field vector of the form
\begin{equation}
\label{eq:bfield}
{\bf B}={\bf B}_{0}\sin{({\bf k}\cdot{\bf x})}
\end{equation}
and is assumed to fill all space. In order to ensure that ${\bf \nabla\cdot B}=0$, we choose 
magnetic-field components of the form \citep{gj94,no06}
\begin{equation}
\label{eq:singlefieldcomp}
\begin{array}{rl}
B_{x} & =B_{0,x}\sin{(k_{y}y+k_{z}z+\sigma_{x})} \\
B_{y} & =B_{0,y}\sin{(k_{x}x+k_{z}z+\sigma_{y})} \\
B_{z} & =B_{0,z}\sin{(k_{x}x+k_{y}y+\sigma_{z})},
\end{array}
\end{equation}
where the $\sigma_i$ terms denote randomly generated phase shifts.
The projections of the wavevector ${\bf k}$ are: 
\begin{equation}
\label{eq:kcomps}
\begin{array}{rl}
k_{x} & =k\sin{\theta}\cos{\eta} \\
k_{y} & =k\sin{\theta}\sin{\eta} \\
k_{z} & =k\cos{\theta}
\end{array}
\end{equation}
where the angles $\theta$ and $\eta$ are randomly selected to generate isotropic turbulence. 
The wavenumber $k$ is randomly generated with uniform distribution in $\ln k$,
such that $2\pi\times 10^{-3}\le k \le 4\pi\times 10^{-1}$ inverse cell 
units. This allows a range of wavelengths between 5 and 1000 cell units, or $20-4000$ periods 
along the line of sight. The amplitude of the magnetic wave follows
\begin{equation}
\label{eq:B0}
B_{0}(k)=B_{0}(k_{min})\left[\frac{k}{k_{min}}\right]^{\frac{1-q}{2}}\quad
k_{min} \le k\le k_{max},
\end{equation}
where we consider two possible values for the power-law index, $q$. For a Kolmogorov 
spectrum $q=5/3$ and for a flat spectrum $q=1$. For a model with a Kolmogorov spectrum,
the rms magnetic amplitude is $\delta B=\sqrt{\langle B_{x'}^2 + B_{y'}^2 +B_{z'}^2\rangle}
\approx 10\mu$G, while a flat-spectrum model
results in $\delta B\approx 20\mu$G.
The flat-spectrum models involve more small-scale fluctuations than do the Kolmogorov models.
We do not know the true shape of the turbulence spectrum in SNRs, but 
comparing the results for the two magnetic-field models permits at least a qualitative 
estimate of the expected results for arbitrary spectra.

The projections of the magnetic-field vector are determined in the same way as for ${\bf k}$:
\begin{equation}
\label{eq:B0comps}
\begin{array}{rl}
B_{0,x} & =B_{0}\sin{\zeta}\cos{\xi} \\
B_{0,y} & =B_{0}\sin{\zeta}\sin{\xi} \\
B_{0,z} & =B_{0}\cos{\zeta},
\end{array}
\end{equation}
except that $\zeta$ and $\xi$ are new angles, still randomly generated such that $-1\le\cos{\zeta}\le 1$ 
and $0\le\xi\le 2\pi$. It is straightforward to develop a turbulent magnetic field from the superposition 
of 1000 waves by summing each component as follows
\begin{equation}
\label{eq:totalfield}
B_{x,tot}=\sum_{i=1}^{1000}{B_{0x,i}\sin{(k_{y,i}\,y+k_{z,i}\,z+\sigma_{x,i})}}.
\end{equation}
While it is important to ensure that ${\bf \nabla\cdot B}=0$ for each wave,
the final magnetic field components must 
also fluctuate along all three coordinate axes, otherwise there would be no field reversals 
along the line of sight and hence no turbulent Faraday rotation. Therefore, we rotate the 
coordinate system by 45 degrees around each axis, thus changing the coordinates 
$(x, y, z)$ to $(x', y', z')$ via
\begin{equation}
\label{eq:newcoords}
{\bf x'}={1\over {2\,\sqrt{2}}}\,\left(\begin{array}{ccc}
1-\sqrt{2} & 1+\sqrt{2 }& -\sqrt{2} \\
-1-\sqrt{2} & -1+\sqrt{2 }& \sqrt{2} \\
\sqrt{2} & \sqrt{2 }& 2 \end{array}\right)\,{\bf x}
\end{equation}
The rows of the transformation matrix give the unit vectors of the primed coordinate system,
${\bf e_{x'}}$, ${\bf e_{y'}}$, and ${\bf e_{z'}}$, and the columns contain the unit vectors 
of the old coordinate system, ${\bf e_{x}}$, ${\bf e_{y}}$, and ${\bf e_{z}}$.
The projections of the magnetic-field vector on the new coordinate axes are
\begin{equation}
\label{eq:newfield}
B_{x'} ={\bf e_{x'}}\cdot \left(B_{x,tot},B_{y,tot},B_{z,tot}\right)
\end{equation}
and analogously for $B_{y'}$ and $B_{z'}$.
Here Eq.~\ref{eq:totalfield} is written in term of the primed coordinates
\begin{equation}
\label{eq:newfield1}
B_{x,tot}=\sum_{i=1}^{1000}{B_{0x,i}\sin{(k_{y,i}\,{\bf e_{y}\cdot x'}
+k_{z,i}\,{\bf e_{z}\cdot x'}+\sigma_{x,i})}}.
\end{equation}
and correspondingly for $B_{y,tot}$ and $B_{z,tot}$. It is straightforward to see that 
$B_{x'}$ now depends on $x'$ etc.

In the primed coordinate system, a three-dimensional grid is defined with size 20x20x20000 
cells or 50x50x20000 cells. The longest dimension (20000 cells) corresponds to the line of 
sight, while the other two dimensions are in the plane of the sky.

\begin{figure}
\plotone{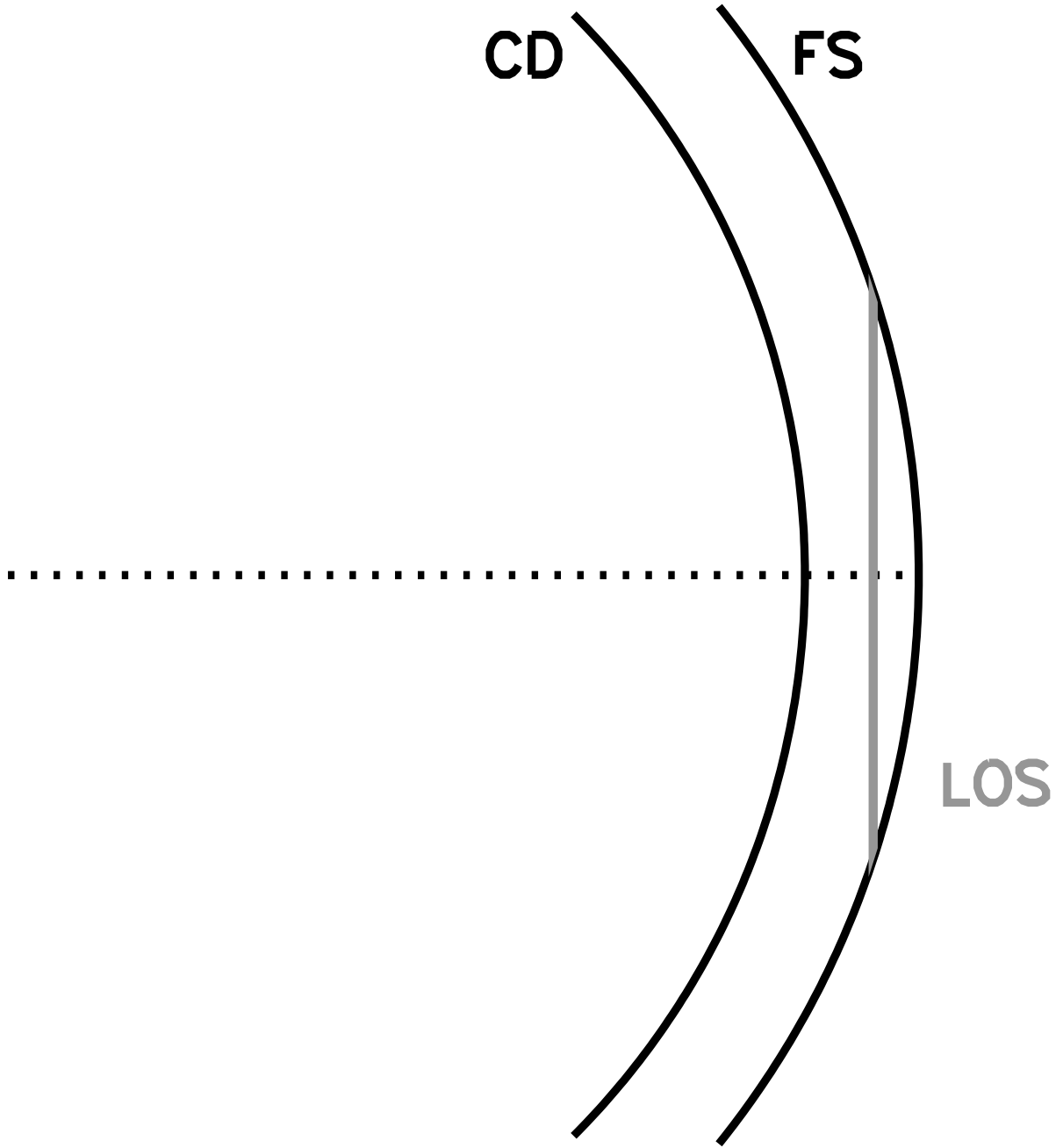}
\caption{Sketch of the geometry. We discuss the transport of polarized radiation for a 
line of sight (LOS) near the rim of the SNR, right behind the forward shock (FS)
but outside the contact discontinuity (CD).
\label{f1}}
\end{figure}
\subsection{Transport of polarized radio emission}\label{rad-trans}
We are interested in modeling the observed polarized radio emission near the rims of the remnant, 
where any signatures of the possibly warped contact discontinuity can be ignored. 
Figure~\ref{f1} shows a sketch of the geometry.
Radio observations of SNRs suggest that the synchrotron intensity, and therefore the 
spectrum of radiating electrons, follows a power law $j_\nu\propto\nu^{-\alpha}$ 
\citep{gr00}.
The emissivity also depends on the perpendicular component of the magnetic field in the emission region: 
\begin{equation}
\label{eq:emissivity}
j_\nu = K\left(\alpha\right)\,B_{\perp}^{\alpha+1}\,\nu^{-\alpha}\qquad{\rm with}\ B_{\perp}^2=B_{x'}^2+B_{y'}^2,
\end{equation}
where $K(\alpha)$ is a constant depending on the density of relativistic electrons \citep{rl79}.
For an isotropic ensemble of electrons the synchrotron emissivity has
a degree of linear polarization 
\begin{equation}
\label{eq:pdeg}
p = \frac{\alpha+1}{\alpha+5/3}
\end{equation}
which is independent of frequency \citep{lr61}. For typical values of $\alpha=0.5$ 
or $\alpha=0.7$ the initial degree of linear polarization is $70\%$ and $72\%$, respectively.

Radiotelescopes typically measure the electric-field direction of a linearly polarized
wave, which reflects the direction of acceleration of the radiating electron. The
\emph{electric}-polarization angle is then rotated by $90^\circ$ to obtain the 
\emph{magnetic}-polarization angle, which reflects the orientation of the perpendicular
component of the magnetic field at the location of the electron. It is therefore practical 
to directly calculate the magnetic-polarization angle, which can be compared with radio
polarimetry data, for which the $90^\circ$-rotation has already been performed.

It is useful to describe polarized emission using a complex polarized intensity 
\citep{burn66}
\begin{equation}
\label{eq:pnu}
P_\nu=I_\nu\,\exp\left(2\,\imath\,\chi\right)\ , 
\end{equation}
the phase of which is twice the magnetic-polarization angle, $\chi$. The factor 2 in the argument 
of the exponential accounts for the indistinguishability of polarization angles 
$\chi$ and $\chi+\pi$. The complex emissivity is then similarly defined as
\begin{equation}
\label{eq:enu}
\epsilon_\nu=p\,j_\nu\,\exp\left(2\,\imath\,\psi\right)\ ,
\end{equation}
where the initial magnetic-polarization angle, $\psi$, reflects the orientation of the 
magnetic field perpendicular to the line of sight through the relations
\begin{equation}
\label{eq:psi}
\cos\psi={{B_{x'}}\over {B_{\perp}}}\qquad \sin\psi={{B_{y'}}\over {B_{\perp}}}. 
\end{equation}
Polarized radio emission will suffer Faraday rotation upon passage through a magnetized plasma. 
The amount of rotation is proportional to the square of the wavelength, and the 
proportionality constant is usually defined as the Faraday 
depth of the source. The final polarization angle 
\begin{equation}
\label{eq:chi}
\chi(\lambda^2)=\psi+\phi\lambda^2,
\end{equation}
where $\phi$ is the Faraday depth calculated as a pathlength integral along the
line of sight of the local electron plasma frequency, $\omega_{p,e}$, the
electron gyrofrequency, $\omega_g$, and the inclination angle between the line of sight
and the magnetic-field vector, $\xi$ \citep{burn66,bb05}.
\begin{equation}
\label{eq:faradep}
\phi(s) =\frac{-1}{4\pi^2\,c^3}\int_0^s\ ds'\,\omega_{p,e}^2\,\omega_g\,\cos\xi\ \nonumber\\ =
\left(0.81\ {\rm rad\, m}^{-2}\right)\,
\int_0^s\,\left(\frac{n_e}{{\rm cm}^{-3}}\right)\left(\frac{B_{\rm los}}{{\rm \mu G}}\right)
\left(\frac{ds'}{{\rm pc}}\right)
\end{equation}
Also, $n_e$ denotes the density of free electrons and $B_{\rm los}=B_{z'}$ is
the magnetic-field component along the line of sight (positive when pointing toward the observer).
In the absence of absorption the transport of polarized radio emission is then given 
by line-of-sight integration.
\begin{eqnarray}
\label{eq:polinten2}
P_\nu & = & \int_{0}^{L_{\rm LOS}} \epsilon_\nu(s)\,
\exp\left[2\imath\,\phi(s)\lambda^2\right] ds \\
& = & \int_{0}^{L_{\rm LOS}} p\ j_\nu(s)\,
\exp\left[2\imath\,\left(\phi(s)\lambda^2 + \psi(s)\right)\right] ds \nonumber
\end{eqnarray}
where $L_{\rm LOS}$ is the length of the line of sight, here 20,000 cells.
We calculate the observed degree of polarization as
\begin{equation}
\label{eq:degpol}
\Pi_\nu = {{\vert P_\nu\vert}\over {I_\nu}}=\frac{\vert\,\int ds\ p\, j_\nu(s) \,
\exp\left[2\imath\,\left(\phi(s)\lambda^2 + \psi(s)\right)\right]\,\vert}{\int ds\ j_\nu (s)}
\end{equation}
and the observed polarization angle according to Eq.~\ref{eq:pnu}. 

It is worth noting three particular aspects of our method.

\begin{enumerate}
\item The degree of polarization does not depend on the
rms amplitude of the perpendicular component of the magnetic field, only on the spatial
fluctuations and the orientation. It also does not depend on the spectral index,
if we ignore the minimal variation of the intrinsic degree of polarization, $p$ (cf.
eq.~\ref{eq:pdeg}).
In the absence of Faraday rotation, i.e. 
if $\vert\phi(s)\vert\lambda^2 \ll\pi$, also the line-of-sight
component of the magnetic field, $B_{\rm los}$, cancels out of Eq.~\ref{eq:degpol}, which is
then solely dependent on the structure of the turbulent field. The frequency cancels as well, 
and therefore we will find that at high frequencies the degree of polarization is inevitably small,
if the magnetic field is fully and isotropically turbulent on small spatial scales. 
A significant homogeneous field or 
anisotropy of the turbulence, e.g. through shock compression, will increase the polarization fraction.
\item Faraday rotation will generally lead to additional depolarization, if
$\vert\phi(s)\vert\lambda^2 \gtrsim \pi$. Our examples assume
numerical values of $n_e=1\,{\rm cm^{-3}}$, 
a line-of-sight length $L_{\rm LOS}=10\,{\rm pc}$, and a mean turbulent magnetic-field strength of 
$\delta B\approx 10\mu$G for the Kolmogorov 
models and $\delta B\approx 20\mu$G for the flat-spectrum models. Variations in one variable 
can be fully compensated by variation in another variable, in particular the wavelength $\lambda$.
Faraday rotation may simply become important at somewhat higher (or smaller) frequency
than shown in our figures. 
Because of the quadratic wavelength dependence, relatively small changes in
wavelength can compensate for relatively large changes in the other variables.
\item We have described the radiation transport along a single line of sight. Since we
simulate a grid of 20x20x20000 cells or 50x50x20000 cells, we can also account for beam
depolarization for a beam size of 20x20 or 50x50 cells, respectively. This beamsize
corresponds to a linear resolution of 0.1\% or 0.25\%
of the line-of-sight length. $L_{\rm LOS}$ is typically a fair fraction of the SNR radius
(see Fig.~\ref{f1}), and hence our beam size corresponds to somewhat less than 0.1\% of the 
angular radius of the SNR. Most published radio observations have larger beams, and
thus we underestimate the beam depolarization.
\end{enumerate}

\section{Results}\label{results}
\subsection{Isotropic strong turbulence}
The most generic case to consider is isotropic turbulence without a large-scale
homogeneous magnetic-field
component. The calculations should broadly describe situations in which $\delta B \gg B_0$,
as is often invoked for young SNRs.
\begin{figure}
\plotone{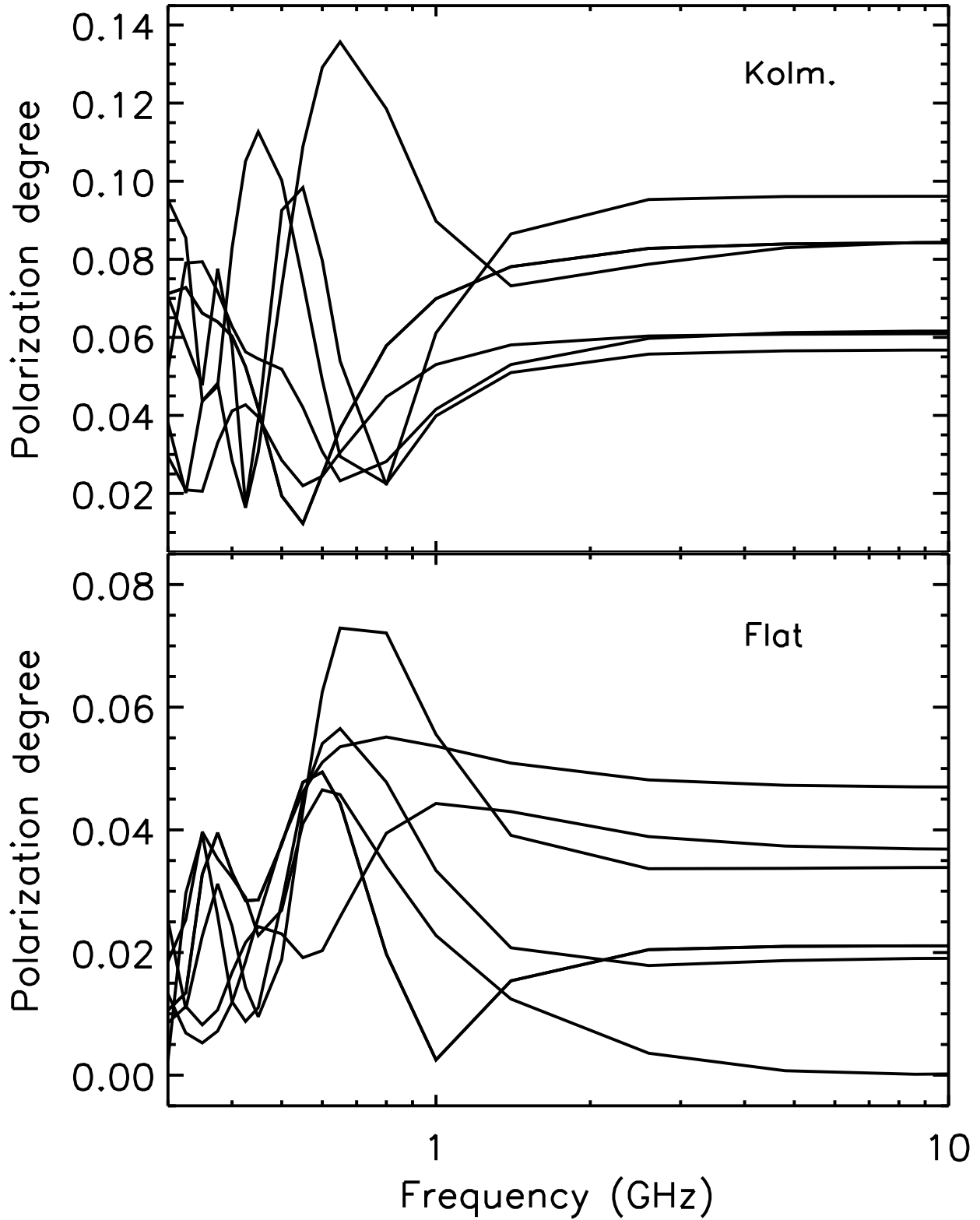}
\caption{The degree of polarization as a function of frequency for a selection of six
turbulence models assuming Kolmogorov spectra (top panel) and six models based on flat spectra
(bottom panel).
\label{f2}}
\end{figure}
Figure~\ref{f2} shows the degree of polarization as a function of frequency for a selection
of magnetic-field models with Kolmogorov and flat turbulence spectra. In all cases the
polarized intensity was "beam-integrated" over an area of 20 by 20 cells.
We find the behavior of
polarized radio-synchrotron emission well characterized by the following statements:
\begin{itemize}
\item Generally, magnetic-field models based on a Kolmogorov turbulence spectrum tend to give
a higher polarization degree than models for flat turbulence spectra. This behavior can
be understood 
with a toy model, in which the magnetic field is assumed constant in a zone of "wavelength"
$l$, but having random variations from zone to zone. If there are $N$ zones on
the line of sight, the mean observed degree of polarization would be about
$70\%/\sqrt{N}$. For the Kolmogorov models we find an
observed degree of polarization around $8\%$, or $N=80$, or $l\simeq 250$~cells.
For flat-spectrum turbulence, which has much higher amplitudes at small scales, we 
find $4\%$, or $N=300$, or $l\simeq 65$~cells. We can extrapolate to turbulence of smaller
wavelength than simulated here, for which we expect the observed degree of polarization to be 
less than $4\%$ on average.
\item The observed polarization angles are uniformly distributed on scales larger than
the characteristic wavelength of the turbulence, which is easily understood in the framework of 
the toy model described under the first bullet above. Because our turbulence model is built 
through the superposition of waves, two lines of sight close to each other will
pass through magnetic-field structures that are somewhat correlated, whereas lines of sight far 
from each other will be fully uncorrelated.
\item Figure~\ref{f3} shows the small-scale 
distribution of the observed magnetic-polarization angle at $4.8$~GHz
for a Kolmogorov model. To be noted from the figure
is that the polarization angle varies by less than $0.3$~radian over an area of 50x50 cells,
whereas it varies wildly over significantly larger areas. Therefore, the beam depolarization
is small, if we beam-integrate the polarized intensity only over
the 50x50 cells, or 20x20 cells as done for Fig.~\ref{f2}. Real
radio measurements have a spatial resolution that is much worse than $0.25\%$ of the
line-of-sight length (50 cells vs. 20,000 cells).
Therefore, beam depolarization may be much more significant in the real data, and the observed
polarization degree may be lower than shown in Fig.~\ref{f2}.
\begin{figure}
\plotone{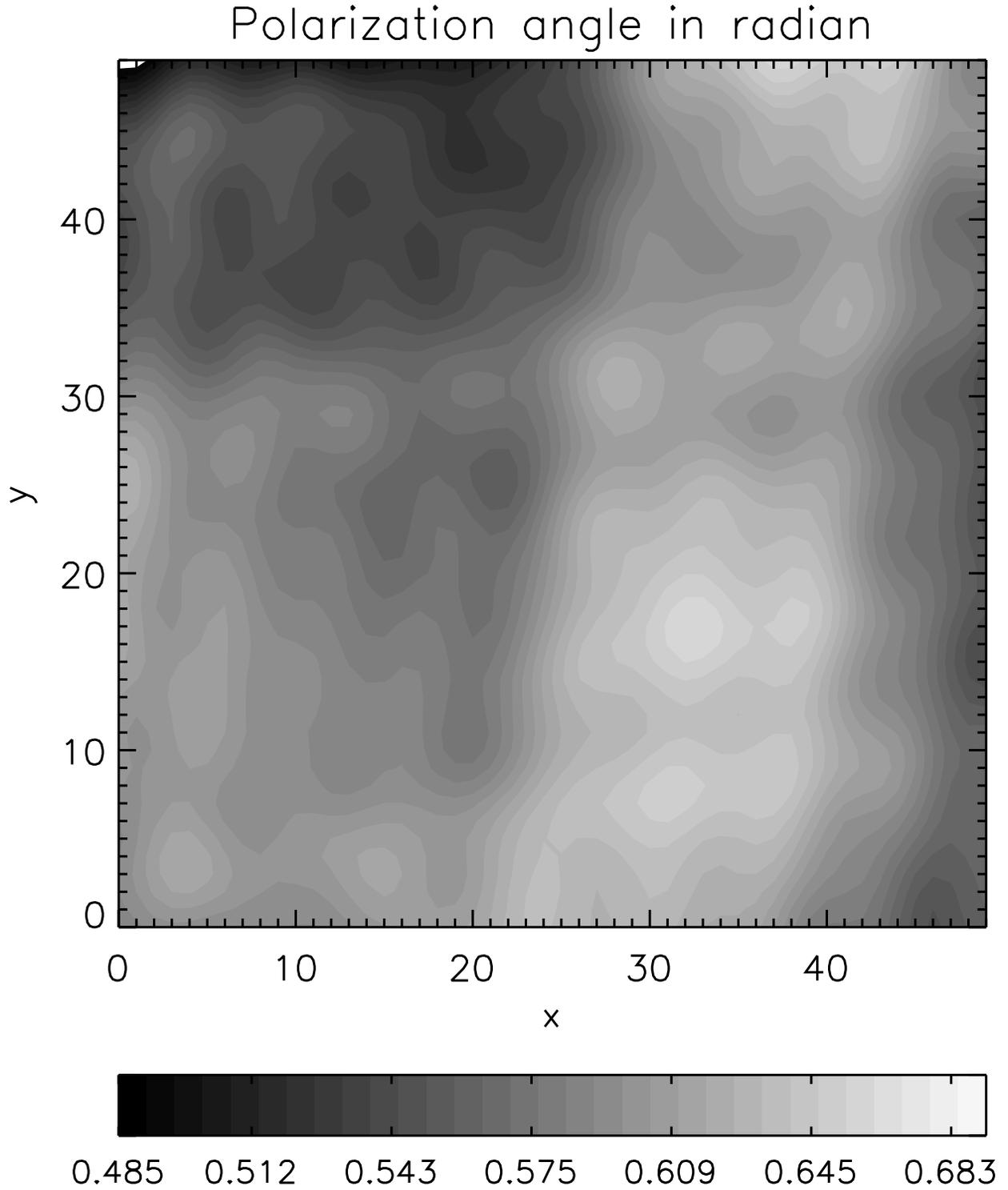}
\caption{The spatial variation of the observed magnetic
polarization angle at $4.8$~GHz, where internal
Faraday rotation is negligible. A turbulence model with Kolmogorov spectrum was chosen.
The plot covers a surface area of 50x50 cells, while the line of sight is 20,000 cells long.
\label{f3}}
\end{figure}
\item At higher frequencies the observed polarization degree is determined only by
depolarization through the intrinsic magnetic-polarization angle, $\psi$, which
reflects the orientation of the perpendicular magnetic-field component along the line-of-sight
(cf. Eq.~\ref{eq:psi}).
\item In Fig.~\ref{f2} we see substantial fluctuations in 
the degree of polarization between different magnetic-field models with the same turbulence spectrum.
We expect that for a wider beam much of this variation would be smoothed out. 
\item Below about $1$~GHz Faraday rotation becomes important. If the product of
rms magnetic-field strength, $\delta B$, free-electron density, $n_e$, and
line-of-sight length, $L_{\rm LOS}$, in any SNR is different from the
canonical values described in Sec.~\ref{rad-trans}, then the onset of
Faraday depolarization may occur at a somewhat lower or higher frequency.
\item Because the magnetic field is turbulent, the effective Faraday rotation in the
observed polarization angle no longer follows a $\lambda^2$-law \citep{bb05}.
Faraday rotation in the foreground -- not considered here -- will follow the
$\lambda^2$-dependency, but it generally cannot be extracted using a $\lambda^2$-law
because it is superimposed on the internal Faraday rotation.
\item The strong Faraday rotation at low frequencies reduces the polarization degree
only weakly, because the polarized emissivities are already fully randomized at
high frequency.
\end{itemize}
\subsection{Mixtures of turbulent and homogeneous fields}
The magnetic field near the rim of SNRs may not be fully turbulent, for example if the
magnetic-field amplification saturates at an amplitude $\delta B \approx B_0$. We can model a
mixture of turbulent and homogeneous fields by adding to our models of fully turbulent
magnetic field a homogeneous field that is oriented in the x-direction.

\begin{figure}
\plotone{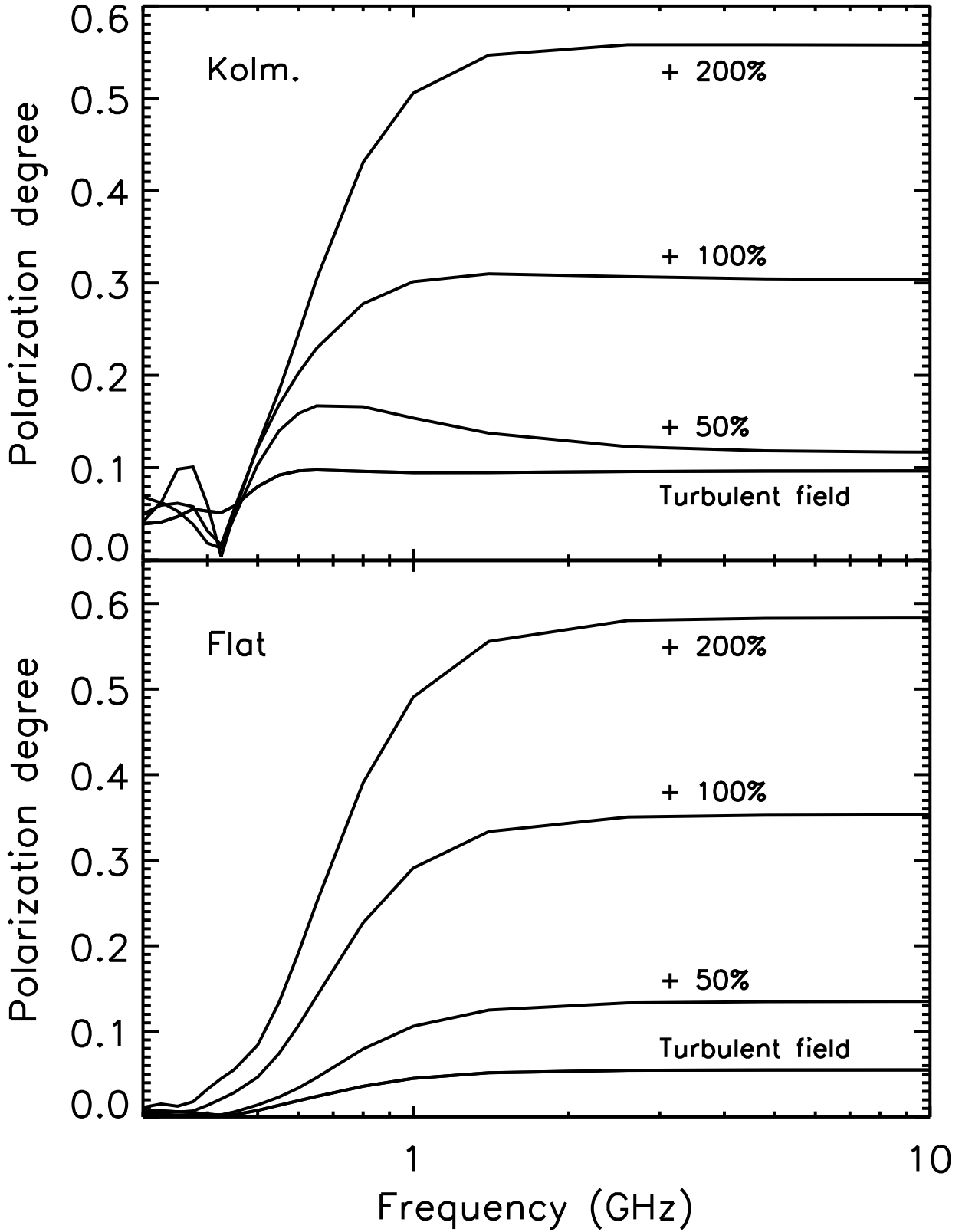}
\caption{The observed polarization degree as a function of frequency for models with a homogeneous
field component of varying strength ($50\%$, $100\%$, or $200\%$ of the rms turbulence amplitude,
$\delta B$). The top panel shows results for Kolmogorov turbulence, and the lower panel displays the
degree of polarization in the case of turbulence with flat spectrum. 
\label{f4}}
\end{figure}

In Fig.~\ref{f4} we show the degree of polarization as a function of frequency for
mixtures of turbulent and homogeneous
fields of varying strength. The polarized intensity has been "beam-averaged" over 50x50 
cells, or 0.25\% of the line-of-sight length, and therefore the curve for a fully
turbulent field is smoother at low frequency compared with those shown in Fig.~\ref{f2}.

Largely independent of the choice of turbulence model, 
a homogeneous magnetic field of the same strength as the rms turbulence amplitude,
$B_0=\delta B$, drives the degree of polarization into the range $30\%$ to $35\%$. A
homogeneous magnetic field at $B_0=2\,\delta B$ gives
a degree of polarization at the level $55\%$ to $60\%$. However, at low frequencies
where internal Faraday rotation is important the degree of polarization remains below
$10\%$, and for flat-spectrum models, i.e. a very small effective
turbulence wavelength, the polarization degree falls below $3\%$.

\begin{table}[tb]
\begin{center}
\begin{tabular}{| l | r | r | r | r |}
\tableline
Model & Turb. only & $+50\%$ & $+100\%$ & $+200\%$ \\ \hline
Flat\_01 & 40.59 & 8.42 & 4.21 & 2.58 \\ \hline
Flat\_02 & 0.43 & 1.92 & 1.20 & 0.81 \\ \hline
Klm\_01 & 35.59 & 22.52 & 5.67 & 1.85 \\ \hline
Klm\_02 & 87.57 & 12.84 & 4.13 & 2.43 \\
\tableline
\end{tabular}
\caption{\label{tab:homofield}Magnetic polarization angles, given in degrees relative to the
direction of the homogeneous field ($B_0\,{\bf e_x}$), for varying ratios $B_0/(\delta B)$ and a
selection of turbulence models.
The frequency is $4.8$~GHz, and so internal Faraday rotation is very small.}
\end{center}
\end{table}
Table~\ref{tab:homofield} gives the observed magnetic-polarization angles at high frequency
for different relative amplitudes of the homogeneous field, which is
oriented in the x-direction (zero angle). Recall that the magnetic-polarization angle differs by 
$90^\circ$ from the electric polarization angle. Already for $B_0=\delta B$
the magnetic-polarization angles reflect the direction of the homogeneous field component 
to within a few degrees. At low frequencies, where internal Faraday rotation is strong,
two lines of sight through statistically independent turbulence give a different
polarization angle as if randomly selected, irrespective of the magnitude of the homogeneous 
field, but the degree of polarization is small. 

\subsection{Shock-compressed turbulence}
If strong magnetic turbulence is generated upstream of the shock, for example by the
streaming of cosmic rays \citep{be04}, then it will
change its properties when transmitted through the shock.
Here we investigate this situation for the simplified case that
the modification by the shock can be fully described by a compression of the spatial scales
in the direction of the shock normal (here the x-axis) and an increase of the perpendicular
magnetic-field components. The change in spatial scale is achieved 
by reallocating the gridpoints of the mesh on which the magnetic field is calculated (see 
sec.~\ref{mf}). Instead of sampling with $\Delta x=\Delta y=\Delta z$ the magnetic-field model
is interpreted as sampled with $\Delta x=(\Delta y)/\kappa$ where $\kappa$ is the compression
ratio. The compression of the perpendicular field components is
effected by the substitution $B_y\longrightarrow \kappa\,B_y$ and likewise for $B_z$.

\begin{figure}
\plotone{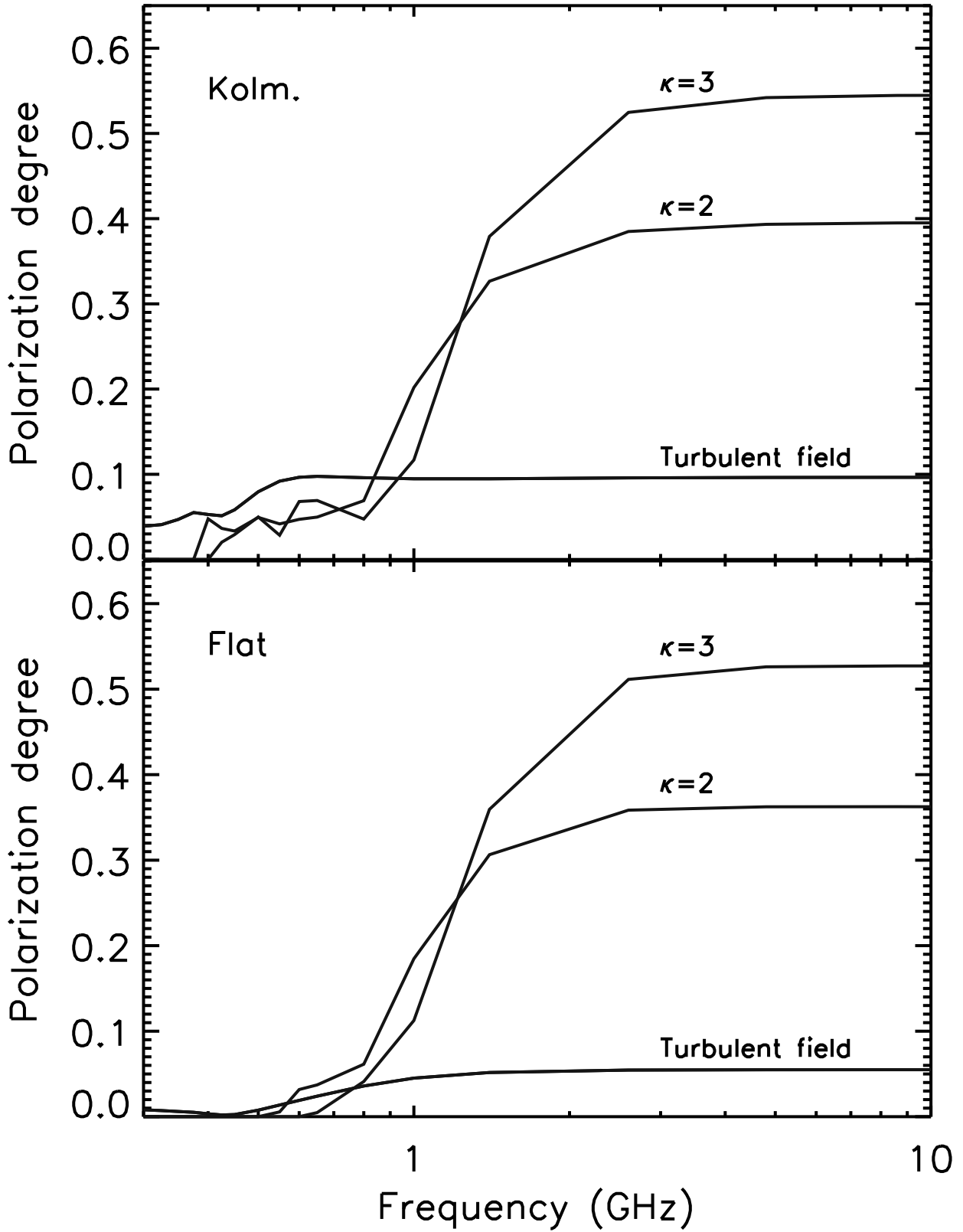}
\caption{The observed polarization degree as a function of frequency for models with
shock-compressed turbulence for a choice of compression ratios, $\kappa$.
The top panel shows results for Kolmogorov turbulence, 
and the lower panel displays the
degree of polarization in the case of turbulence with flat spectrum.
\label{f5}}
\end{figure}

In Fig.~\ref{f5} we show the polarization degree as a function of frequency for
shock compression with $\kappa=2$ and $\kappa=3$, where the "beam-averaging"
is effectively over 20x20 and 15x15 cells, respectively, so the "beamsize" decreases with
increasing $\kappa$. Cosmic rays can modify the shocks at
which they are accelerated, resulting in a reduction of the
compression ratio at the subshock of the thermal gas and an increase in the 
overall compression \citep{rey08}. The choice of compression ratios in figure~\ref{f5}
is solely meant to illustrate the behavior of the polarization, and in fact the changes
in the degree of polarization will be only more pronounced for larger compression ratios.
To be noted from the figure is that shock compression
will substantially increase the degree of polarization at higher frequencies:
already for a compression ratio
$\kappa=2$ the degree of polarization falls into the range $35\%$ to $40\%$,
largely independent of the choice of turbulence model. As in the case of a mixture with
a homogeneous field, Faraday rotation will strongly depolarize the emission
at low frequency. The onset of efficient internal Faraday rotation is observed 
at somewhat higher frequency than in the case of a homogeneous field component perpendicular
to the line of sight, because the shock compression also increases the line-of-sight
component of the magnetic field and hence the Faraday depth.
While shock compression or a homogeneous field component can increase the degree of 
polarization, they will inevitably also modify the observed magnetic-polarization angle
which is shown in Table~\ref{tab:compshocks}. Even for 
a moderate compression ratio $\kappa=2$, at high radio frequencies
the magnetic-polarization angles lie around 
$90^\circ$ to the shock normal with very little scatter (the electric polarization is
along the shock normal). At low frequencies Faraday rotation 
randomizes the polarization angle, but the degree of polarization is generally small. 
We can conclude that shock compression will not give a high degree of polarization 
$\gtrsim 20\%$ \emph{and} a magnetic polarization aligned with the shock normal.

\begin{table}[tb]
\begin{center}
\begin{tabular}{| l | r | r | r |}
\hline
Model & Turb. only & $\kappa=2$ & $\kappa=3$ \\ \hline
Flat\_01 & 40.59 & 91.42 & 93.48 \\ \hline
Flat\_02 & 0.43 & 91.84 & 92.63 \\ \hline
Klm\_01 & 35.59 & 85.66 & 89.46 \\ \hline
Klm\_02 & 87.57 & 92.48 & 93.72 \\
\tableline
\end{tabular}
\caption{\label{tab:compshocks}Magnetic polarization angles, given in degrees relative to the
shock normal, for varying compression ratios and a selection of turbulence models.
The frequency is $4.8$~GHz, and so internal Faraday rotation is very small.}
\end{center}
\end{table}
\section{Discussion and summary}
We have seen that isotropic strong turbulence will produce weakly polarized radio emission
even in the absence of internal Faraday rotation. The polarization angle
can be expected to vary 
on spatial scales of the order of the typical wavelength of the magnetic-field fluctuations,
and therefore radio polarimetry data of very high angular resolution are needed to
observe those variations.

If anisotropy is imposed on the magnetic-field structure, the degree of polarization
can be significantly increased, provided internal Faraday rotation is inefficient. Both
for shock compression and a mixture with a homogeneous field the increase in polarization degree
goes along with an alignment of the observed magnetic-polarization angle with the direction of
the dominant magnetic-field components. In the case of shock compression we therefore expect
tangential magnetic polarization at the rims of SNRs, or electric polarization vectors that
are predominantly radial in orientation.

Few young SNRs are suitable for the study of strong magnetic turbulence through
radio polarimetry. Some remnants like Cas~A feature ejecta beyond
the nominal forward shock \citep{hwa04},
and therefore our assumption of a simple, spherically symmetric structure does not apply.
Also, the line-of-sight should not pass through the contact discontinuity, at which the
magnetic field may be very strong \citep{Rosenau76,lp04}. It is well known that the
contact discontinuity is hydrodynamically unstable \citep[e.g.][]{be01}, and so in projection
it will appear as an extended feature with a turbulent field structure. Also, if the SNR efficiently
accelerates cosmic rays, the contact discontinuity will be
closer to the forward shock than in a purely hydrodynamical SNR. Using
X-ray measurements, \citet{warren} on average find traces of the contact discontinuity in Tycho's SNR 
out to 93\% of the projected radius of the forward shock.
\citet{cc08} use the same technique on data of the remnant of SN~1006 and find that in the regions
of bright non-thermal X-ray emission the contact discontinuity extends all the way to the
forward shock. In both remnants the proximity of forward shock and the contact discontinuity  
presumably arises from a combination of hydrodynamical instabilities of the contact discontinuity and 
structural modifications on account of cosmic-ray acceleration.
In any case, in SN~1006 we may not find a line of sight that is clearly inside the forward shock but
outside the contact discontinuity. For Tycho's SNR 
radio-polarimetry data are required with an angular resolution around 1\% of the angular radius, so 
the forward shock and the contact discontinuity are at least a few beamwidth apart.

\citet{del02} observed the polarized synchrotron emission from Kepler's SNR at 6~cm and
20~cm wavelength. After rotating the measured electric polarization by $90^\circ$, 
they found predominantly radial magnetic polarization in the outer regions of the remnant,
where the degree of polarization was a few per cent at 6~cm and less than that at 20~cm.
The angular resolution is moderate, though, and the beam size is about 7\% of the projected radius
of the forward shock. This is a factor of 100 worse than the beamsize assumed for the figures in
this paper, and a direct comparison is therefore difficult. Also, the contact discontinuity and the 
forward shock are not clearly separated at this resolution, and so any inferred
field orientation and turbulence level cannot be unambiguously associated with the magnetic-field
structure directly behind the forward shock.

\citet{di91} presented radio polarimetry data of Tycho's SNR at 6~cm and 20~cm wavelength.
The spatial resolution is about $0.7\%$ of the forward-shock radius, roughly comparable to 300 cells
in our magnetic-field model. The radio morphology can be well described by an outer
rim and an inner shell of outer radius 0.92 SNR radii \citep{ks00}. The outer rim is positionally
coincident with the outer periphery of the X-ray emission, suggesting that the inner shell marks the 
location of the contact discontinuity and the reverse shock. We conclude that our model
is applicable to polarized radio emission from the outer rim.

In the outer rims, the percentage polarization at 6~cm wavelength is typically in the range
$20\%$ to $30\%$, and the field orientation is radial \citep[cf. Fig.5a of][]{di91}.
Beginning about 10\arcsec (or $7\%$ of the SNR radius) toward the interior of the remnant,
the degree of polarization
is lower and the polarization angle varies on scales of about 10\arcsec, consistent with
large scale turbulence near the contact discontinuity. Internal Faraday rotation is likely
negligible, because no deviation from a $\lambda^2$-law is observed between 6~cm and 20~cm.
The lack of efficient internal Faraday rotation can be used to set limits on the amplitude of
the turbulent magnetic field. The post-shock gas density in Tycho is not well known, but the estimates
include our canonical number $n_e=1\ {\rm cm^{-3}}$. The distance is likely in the range 1.5~kpc
to 3~kpc \citep{sm91}, and therefore the line-of-sight length at the rim is $L_{\rm LOS}\approx
0.5$~pc. For the canonical numbers used in our model internal Faraday rotation becomes important 
at about 1~GHz, which with the gas density and physical size of Tycho's SNR requires
$\delta B\simeq 200\ {\rm \mu G}$. If the field strength were significantly larger than that,
\citet{di91} should have observed internal Faraday rotation.

Comparing with our model results presented in Sec.~\ref{results}, we find no evidence for isotropic
or shock-compressed magnetic turbulence that is amplified to an amplitude larger than the homogeneous 
magnetic field. The radio data are compatible with two scenarios:
a turbulent magnetic field superimposed on a
radial large-scale field of similar amplitude, i.e. a parallel forward shock with
$\delta B\simeq B_0$, or, alternatively, anisotropic turbulence that has a somewhat larger 
magnetic-field amplitude in the radial direction. 

Anisotropic magnetic turbulence with preferentially radial orientation is not expected in models 
involving magnetic-field amplification by streaming cosmic rays in the upstream region, in particular 
on account of shock compression. We note with interest that recent MHD simulations
of magnetic-field growth following shock distortions on account of density fluctuations in
the upstream medium suggest that the radial magnetic-field components may be slightly stronger
than those in the shock plane \citep{zp08}. Additional simulations with higher resolution 
are needed to confirm these findings and establish their dependence on the wavelength of the
density fluctuations and other parameters \citep[e.g.][]{cho}.

\acknowledgements
We acknowledge support by NASA under award No. NAG5-13559.

\end{document}